\documentclass[
superscriptaddress,
 amsmath,amssymb,
 aps,
prb,
reprint
]{revtex4-2}
\usepackage[utf8]{inputenc}
\usepackage{wrapfig}
\usepackage{xcolor}
\usepackage{graphicx}
\usepackage{dcolumn}
\usepackage{bm}
\usepackage{epstopdf}
\usepackage{gensymb}
\usepackage{hyperref}
\usepackage{natbib}
\usepackage{float}
\usepackage[T1]{fontenc}

\begin{document}

\title{Adsorption-Controlled Epitaxy and Twin Control of $\gamma$-GaSe on GaAs (111)B}

\author{Joshua Eickhoff}
\affiliation{University of Wisconsin--Madison, Department of Materials Science and Engineering}
\affiliation{U.S. Army DEVCOM Army Research Laboratory, Adelphi, MD 20783}

\author{Wendy L. Sarney}
\affiliation{U.S. Army DEVCOM Army Research Laboratory, Adelphi, MD 20783}

\author{Sina Najmaei}
\affiliation{U.S. Army DEVCOM Army Research Laboratory, Adelphi, MD 20783}

\author{Daniel A. Rhodes}
\affiliation{University of Wisconsin--Madison, Department of Materials Science and Engineering}
\affiliation{University of Wisconsin--Madison, Department of Physics}

\author{Jason Kawasaki}
\affiliation{University of Wisconsin--Madison, Department of Materials Science and Engineering}
\affiliation{University of Wisconsin--Madison, Department of Physics}
 
\date{\today}

\begin{abstract} 

The III-Se layered semiconductors, including InSe and GaSe, are promising optoelectronic materials due to their relatively high electron mobilities at room temperature, nonlinear optical responses, ferroelectricity, self-passivated van der Waals surfaces, and ability to alloy and synthesize heterostructures for bandgap engineering. 
Adsorption control is a widely utilized strategy for controlling the stoichiometry and phase formation of these materials; however, the bounds of the adsorption-controlled growth window for GaSe have not been systematically established. Additionally, challenges with control over polytype and twinning remain.
Here, we use molecular beam epitaxy to experimentally map the adsorption-controlled growth window of GaSe films on vicinal GaAs (111)B substrates. The observed phase boundaries show qualitative agreement with Ellingham diagram predictions. All films crystallize in the $\gamma$ ($R3m$) polytype. 
Increasing growth and annealing temperature leads to decreased mosaicity measured by x-ray rocking curve and smoother surfaces measured by atomic force microsocopy, at the expense of a transition from singly oriented $\gamma$ to twinned $\gamma$ with $60\degree$ rotated domains.

\end{abstract}

\maketitle

\section{Introduction}

The III-Se post transition metal monochalcogenides (e.g., GaSe, InSe, and their alloys) host attractive properties for optoelectronic and spintronic applications, including a thickness tuned parabolic to M-shaped dispersion of the valence bands~\cite{Jung2015, Budweg2019}, van der Waals interfaces that relax the lattice matching constraints for heteroepitaxy \cite{Claro2023}, and the ability to engineer band gaps through alloying  \cite{tran2024molecular} or synthesis of heterostructures~\cite{Claro2023}.
These materials crystallize in tetralayer Se-III-III-Se blocks separated by van der Waals gaps, where differences in tetralayer-tetralayer stacking registry define the different polytypes.
Compared to transition metal dichalcogenides (e.g., MoSe$_2$, WSe$_2$, MoS$_2$) that contain low vapor pressure refractory metals like Mo or W, the higher vapor pressure of group III metals makes the III-Se materials more amenable to large-area scalable synthesis by physical vapor deposition techniques, including molecular beam epitaxy (MBE) \cite{kojima1994epitaxial,Rajan2017, Law2024, jedrecy1997epitaxy, eddrief1994heteroepitaxy}. However, key synthesis challenges for these materials remain, including stoichiometry control, polytype control, and suppression of twinning.

While adsorption-control is often implicitly utilized to control the stoichiometry of III-Se films during MBE growth \cite{kojima1994epitaxial,Rajan2017, Law2024, jedrecy1997epitaxy, eddrief1994heteroepitaxy}, for GaSe the bounds of this window and its thermodynamic basis have not been systematically established. Adsorption control, the synthesis of a line compound that is in equilibrium with a volatile constituent, is widely utilized in MBE growth \cite{cho1975molecular, arthur1968interaction, du2020control, shourov2020semi, jalan2009molecular, ihlefeld2007adsorption}. The canonical example is GaAs, which has self-limited stoichiometry over a wide range of arsenic overpressure \cite{cho1975molecular, arthur1968interaction}. This self-regulated stoichiometry, and the associated point defect concentrations below parts per billion, has enabled world record electron mobilities in MBE-grown GaAs \cite{pfeiffer1989electron, tsui1982two} with applications for high electron mobility transistors, lasers \cite{norman2019review}, and quantum sciences \cite{nakamura2019aharonov}. 
A recent study mapped the stability of the line compound InSe versus volatile Se and competing In-Se phases \cite{liu2023growth}; a similar relationship is expected for GaSe but has not yet been systematically mapped and compared to theory.

\begin{figure}[h]
    \centering
    \includegraphics[width=0.5\textwidth]{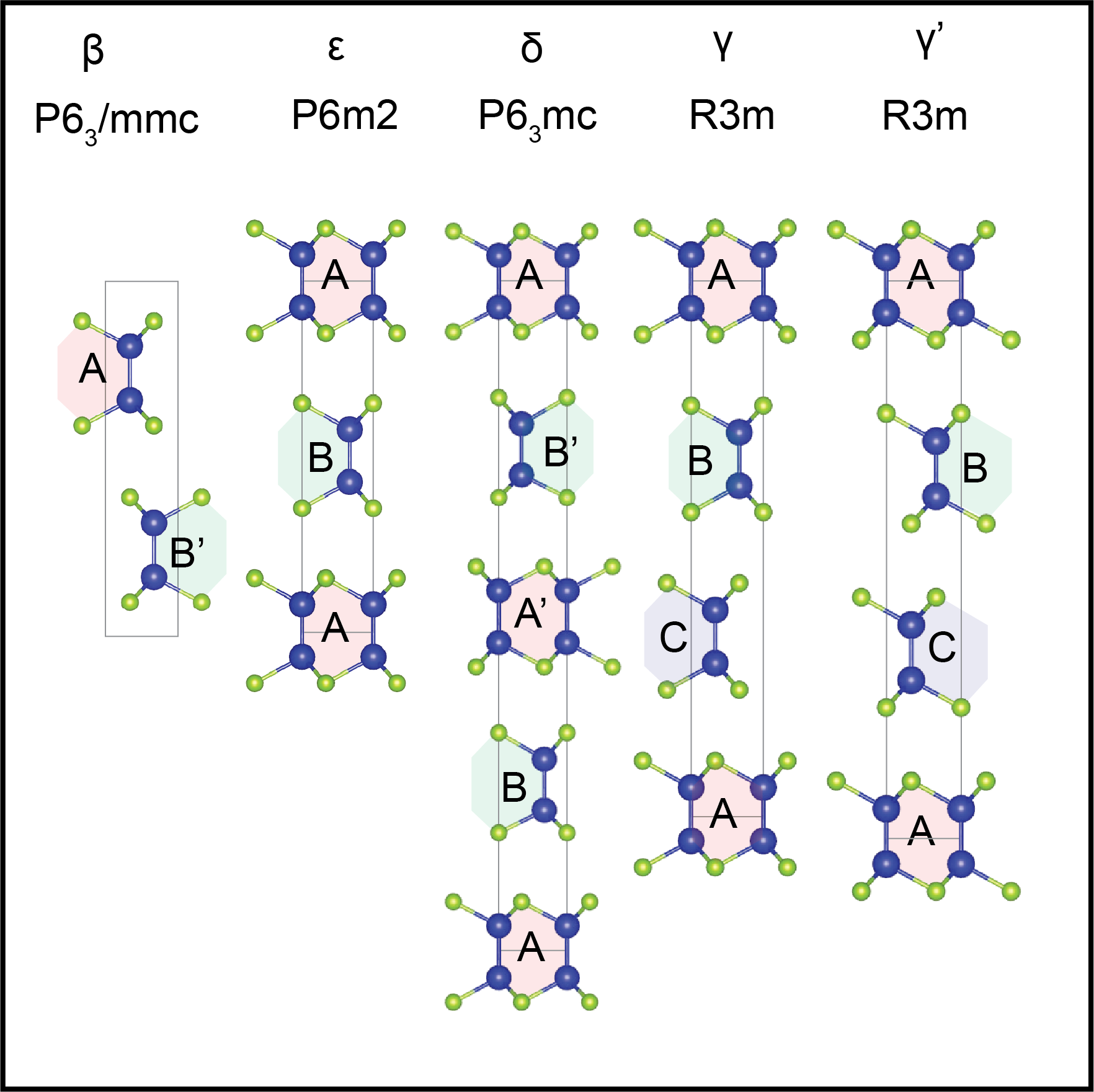}
    \caption{Common GaSe polytypes.}
    \label{fig:stacking}
\end{figure}

A second challenge is control of polytypes and twinning. GaSe crystallizes in $\beta$ (2R), $\epsilon$ (2H), $\gamma$ (3R), $\gamma'$ (3R), and $\delta$ (4H)  polytypes \cite{grzonka2021novel, jedrecy1997epitaxy, sorokin2019molecular, shiffa2024wafer, Law2024, grzonka2021novel, yonezawa2019atomistic, Rajan2017, diep2019screw} (Appendix Figure \ref{fig:stacking}), each with similar formation energies \cite{Law2024,Dong2020}. The most commonly observed polytype for MBE grown GaSe is $\gamma$ ($R3m$), which typically forms 60$\degree$ rotated twins ~\cite{grzonka2021novel, shiffa2024wafer, Law2024, yonezawa2019atomistic}. Controlling twins is essential for electronic and optoelectronic devices because twin boundaries can act as non-radiative recombination centers~\cite{Mathew2025Recomb} and can substantially scatter electrons, reducing carrier mobilities \cite{park2019electron}. 

\begin{figure*}
    \centering
    \includegraphics[width=0.95\linewidth]{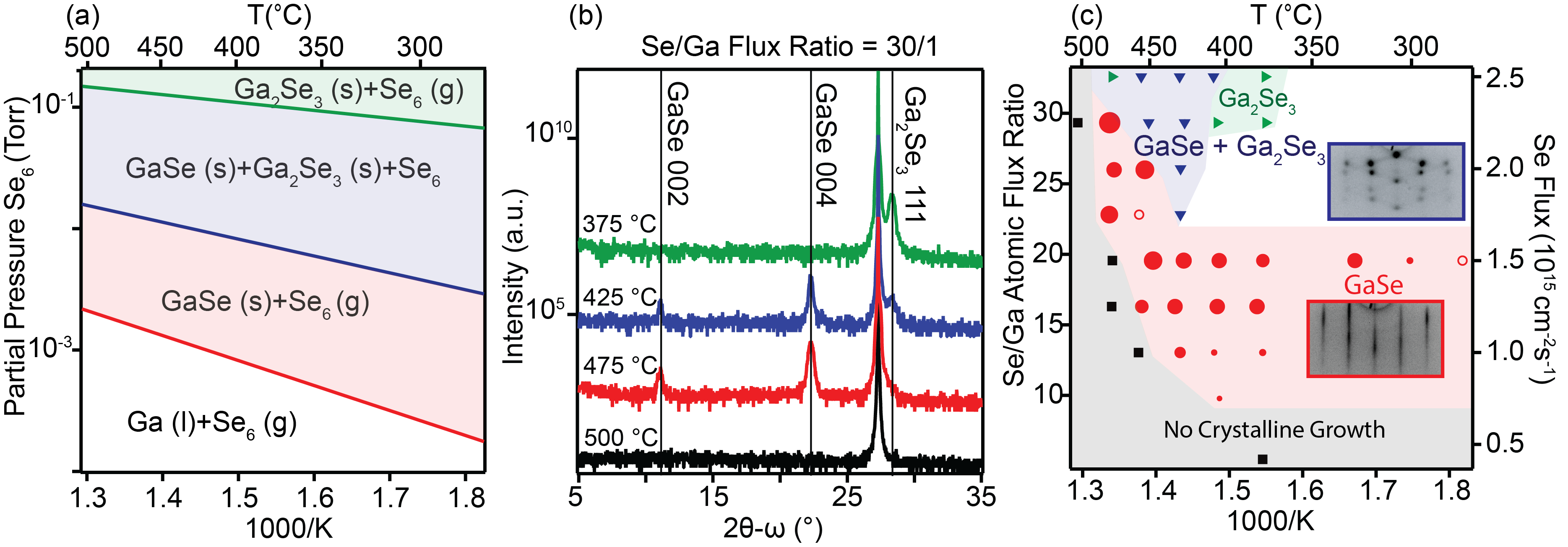}
\caption{(a) Ellingham diagram for the Ga-Se calculated using Se$_6$ as the gaseous species. 
(b) Symmetric $2\theta-\omega$ x-ray diffraction scans for films grown at a Se/Ga atomic flux ratio of 30 for varying growth temperatures. 
(c) Experimentally observed phases for MBE growth of GaSe on GaAs (111)B. Inserts show Reflection High Energy Electron Diffraction (RHEED) patterns for GaSe and mixed GaSe + Ga$_2$Se$_3$. Filled marker size corresponds to the inverse of the GaSe 004 $\omega$ rocking curve full width at half maximum ($1/\Delta\omega_\mathrm{fwhm}$), which scales from 1/1.172$\degree$ to 1/4.506$\degree$. Open marker indicates no rocking curve measured.}
\label{fig:phases}
\end{figure*}

Here, we experimentally map the adsorption-controlled growth window for GaSe films by MBE on GaAs (111)B substrates. The observed phase boundaries qualitatively agree with the thermodynamic prediction from an Ellingham diagram. We find that growth temperature produces a tradeoff of crystal quality metrics: lower temperature growth produces simply oriented $\gamma$ with rough surfaces and broad X-ray rocking curve widths, whereas high temperature improves the roughness and rocking curve width at the expense of forming 60$\degree$ twins. We speculate that overcoming kinetic barriers may be responsible for twinning at high temperatures.

\section{Methods}
\label{sec:methods}

GaSe films were grown in a homemade chalcogenide MBE system. 
GaSe films were grown on epi-ready GaAs (111)B with a $4\degree$ miscut towards $\langle 110 \rangle $ (AXT). The in-plane lattice parameters of $a_{GaSe}=3.74$ \AA\ and $a_{GaAs,110}=3.995$ \AA\ correspond to a $-6.5\%$ mismatch. Substrates were indium bonded to sample plates and pumped in a high vacuum ($10^{-8}$ Torr) loadlock before loading into the growth chamber. In the MBE chamber, substrates were heated to 580 $\degree$C with a Se flux of $\sim 10^{15}$ atoms/cm$^2$s for 20 minutes to desorb surface oxides, as monitored by the reflection high energy electron diffraction (RHEED). Sample temperatures were then decreased to $300-500 \degree$C for GaSe film growth. Sample temperatures were measured using a thermocouple that was calibrated to the native oxide desorption temperature of GaAs and to the melting point of indium.

GaSe films with a nominal thickness of 45 nm were grown via co-deposition using a standard effusion cell for Ga ($9.99999\%$) and a hot lipped effusion cell for Se ($99.999\%$). The hot lip is operated at 250 $\degree$C to produce a mixture of Se$_2$, Se$_5$, Se$_6$, and Se$_7$, where the resulting Se flux is primarily Se$_6$~\cite{cheng1990molecular}. The Ga flux was fixed at at $7.67 \times 10^{13}$ atoms/cm$^2$s, while the Se flux was varied from $5 \times 10^{14}$ to $2.5 \times 10^{15}$ atoms/cm$^2$s. Effusion cell fluxes were measured using a quartz crystal microbalance  calibrated by Rutherford Backscattering Spectrometry. 

\begin{figure}
    \centering
    \includegraphics[width=0.5\textwidth]{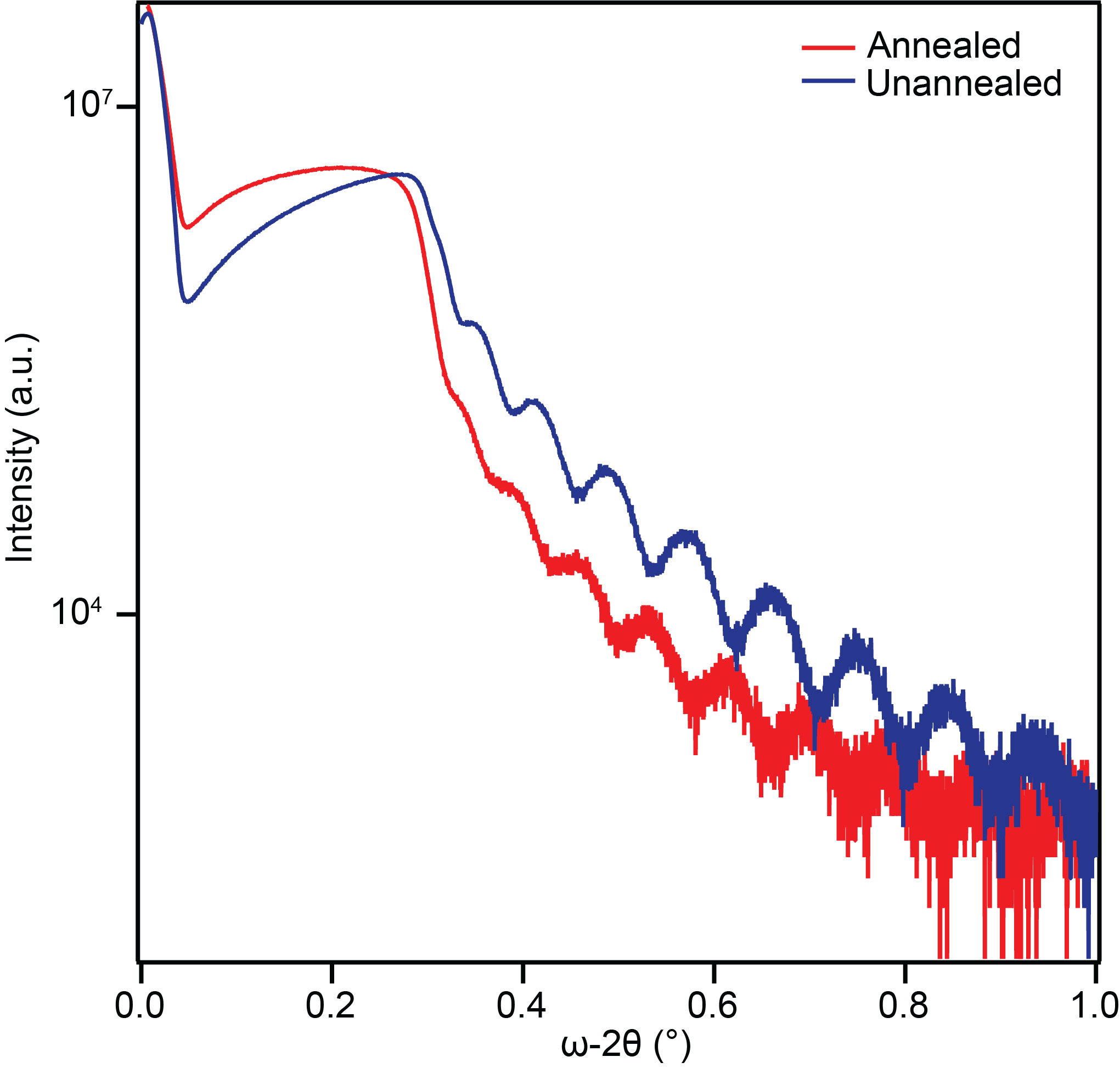}
    \caption{Impact of high temperature anneal on film thickness. X-ray reflectivity measurements comparing GaSe film grown at 400 $\degree$C (blue, measured thickness 43 nm), with a film grown at 400 $\degree$C for the same growth time and then annealed to 520 $\degree$C (red, measured thickness 48 nm).  }
    \label{fig:XRR}
\end{figure}

Several GaSe films were annealed up to $520 \degree$C after film growth to increase the atomic scale ordering. Previous studies demonstrate that while the Ga-Se sticking coefficients are vanishingly small for film growth above $500 \degree$C ~\cite{Law2024,Yu2023,Kis2018}, GaSe films can be annealed to higher temperatures after growth using appropriately matched Ga and Se fluxes to compensate for film desorption from the surface. To perform these anneals, we first grew GaSe at $400 \degree$C and then shuttered the Ga. The sample temperature was then increased to 520 $\degree$C at a rate of 20 $\degree$C/min. Subsequently, both Se and Ga shutters were opened and the sample was held at At $520 \degree$C for an additional 30 minutes. Then both shutters were closed and the samples were cooled to room temperature at a rate of $20 \degree$C/min in a residual background of Se~\cite{Kis2018}. X-ray reflectivity (XRR) measurements (Appendix Figure \ref{fig:XRR}) show that this anneal procedure results in a thickness change from 43 nm before the anneal to 48 nm after the anneal.

2$\theta$-$\omega$ survey scans, $\omega$ rocking curves, and in-plane azimuthal $\varphi$ scans were measured using a Panalytical Empyrean X-ray diffractometer using Cu K$\alpha$ radiation with a 4 bounce Ge 220 monochromater on the incident side. No monochromater was used on the analyzer side. AFM images were acquired in tapping mode using an Asylum Cypher S atomic force microscope (Asylum Research/Oxford Instruments, USA) with a silicon tapping mode cantilever (AC160TSA R3; Olympus/Asylum Research).

\begin{figure}
    \centering
    \includegraphics[width=1\linewidth]{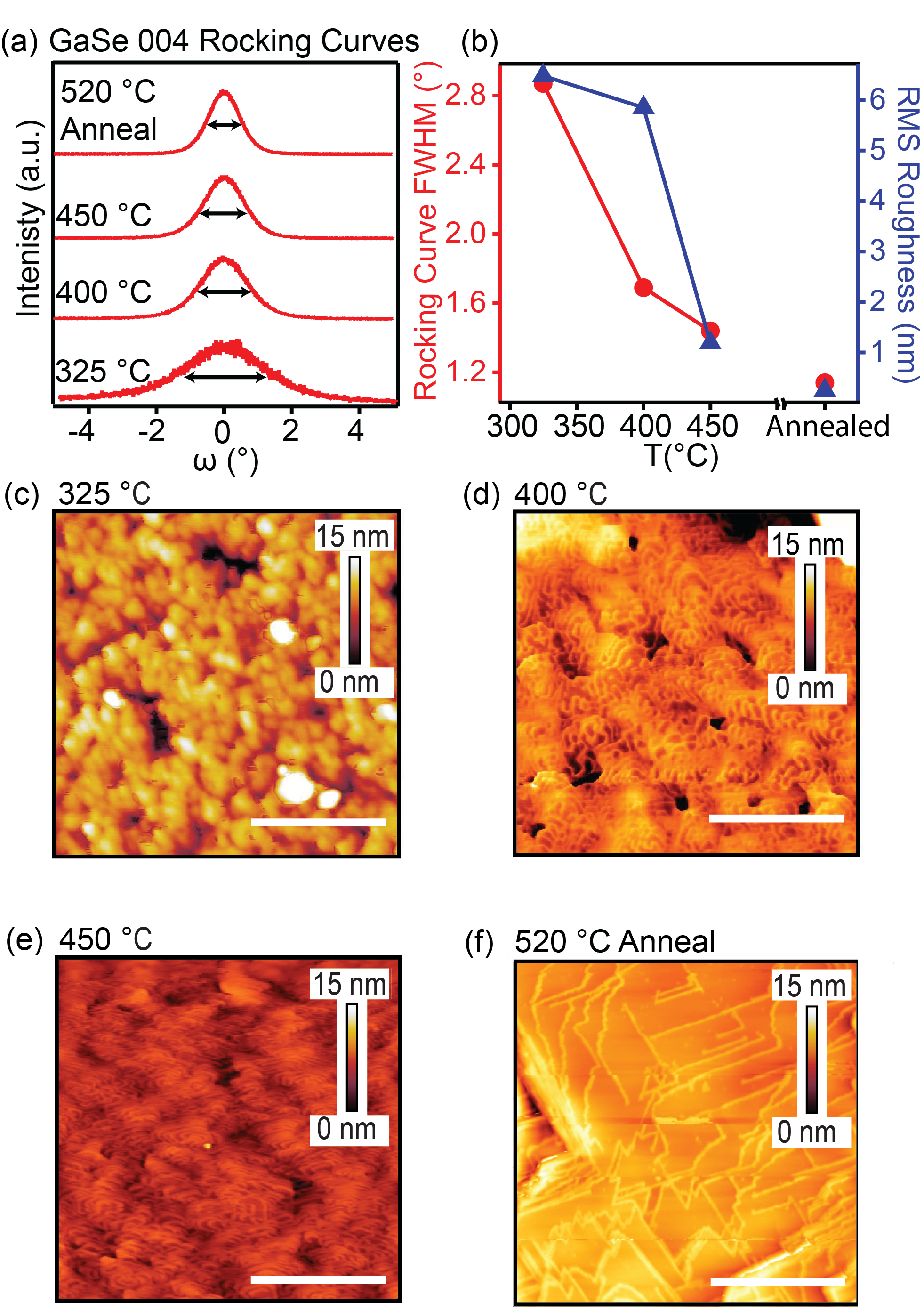}
       \caption{(a) X-ray rocking curves of the GaSe 004 reflection for fixed Se/Ga flux ratio 20/1. (b) Summarized rocking curve width and root mean square (RMS) roughness versus growth or anneal temperature.(c-f) Atomic Force Microscopy (AFM) topography images of GaSe samples grown at 325 $\degree$C, 400 $\degree$C, 450 $\degree$C, and 400 $\degree$C plus a post growth anneal at 520 $\degree$C. Scale bars are 400 nm.}
    \label{fig:temperature}
\end{figure}
	
Cross-sectional TEM samples were prepared with a Thermo Scientific Helios G4 UX Dualbeam focused ion beam (FIB) system. Selected area diffraction patterns (SAED, aperture size 20 $\mu$m) and high angle annular dark field (HAADF) images were collected from an aberration-corrected JEOL ARM 200F scanning transmission electron microscope (STEM) operated at 200 keV. Based on the dimensions of the electron transparent regions of the TEM specimens, the SAED patterns typically probed a total area of 40 $\mu$m$^2$ which includes both GaSe film and GaAs substrate.
We use linear transformations to align the selected area diffraction patterns to crystallographic directions and convert from camera pixel spacing to physical units of $Q$. These transformations included brightness scaling, translation, rotation, and linear $x,y$ scaling, aligning to the GaAs substrate $22\bar{4}$ reflections. In Appendix Figure \ref{fig:Adjustment} we show an example of this rescaling for the film grown at 400 $\degree$C.
Simulated x-ray diffraction patterns were computed using CrystalMaker SingleCrystal based on $\beta$, $\delta$, $\gamma$, and $\epsilon$-GaSe structure files from the Inorganic Crystal Structure Database (ICSD) ~\cite{Kuhn1975DeltaGaSe, Benazeth1988BetaGaSe, Cenzual1991EpsilonGaSe, Schubert1955GammaGaSe,Stevenson1994GaAs}. For $\gamma'$, which does not appear in ICSD, we use the atomic positions reported in Ref. ~\cite{Law2024}. 

\section{Results and Discussion}
To guide synthesis we begin by computing the Ellingham diagram, which describes the bulk thermodynamic stability of Ga-Se phases as a function of temperature, $T$, and Se partial pressure. Note this should be considered a qualitative rather than quantitative guide, since the bulk equilibrium Ellingham diagram does not consider contributions from surface energies, strain, and kinetic factors.
We construct this diagram based on tabulated bulk thermodynamic data \cite{Berkowitz1966,Gaur1981,Ider2015,Tyurin2005,SEDMIDUBSKY2019}. For simplicity we compute the diagram for Se$_6$, which is the primary volatile species expected from our hot lipped Se effusion cell operated at 250 $\degree$C~\cite{Yamdagni_1968}. In Appendix Figure \ref{fig:ellingham}, we compare with a similar calculation for Se$_2$. 

In Fig. \ref{fig:phases}(a), the red shaded region bounded between the red and blue lines defines the expected adsorption-controlled growth window for GaSe. In between these lines, solid GaSe exists in equilibrium with Se vapor, and since GaSe is a line compound its stoichiometry is self-limited. The lower red line is derived from the Gibbs free energy for the GaSe decomposition reaction GaSe(s) $\Leftrightarrow$ Ga (l) + $\frac{1}{6}$ Se$_6$ (g). Below this curve, solid GaSe decomposes into liquid Ga and vapor Se. 
The blue curve is defined by the reaction Ga$_2$Se$_3$ $\Leftrightarrow$ 2 Ga (l) + ½ Se$_6$ (g). Above this line, Ga$_2$Se$_3$ becomes stable along with GaSe, leading to a region where both GaSe and Ga$_2$Se$_3$ are stable (blue region). The green line is defined by the reaction 4GaSe (s) $+$ Se$_6$ (g) $\Leftrightarrow$ 2Ga$_2$Se$_3$ (s) $+$ 2Se$_2$ (g). Pure Ga$_2$Se$_3$ and Se vapor exist above this line.

To experimentally validate this prediction, we synthesize films by MBE on intentionally 4$\degree$ miscut, As-terminated GaAs (111)B substrates over a range of growth temperatures and relative Se fluxes (Methods). 
Figs \ref{fig:phases}(b,c) present representative reflection high energy electron diffraction (RHEED) patterns and symmetric $2\theta-\omega$ x-ray diffraction (XRD) measurements of GaSe films. For decreasing growth temperature at fixed Se/Ga flux, we observe an evolution from no crystalline growth (black curve), to phase pure GaSe (red), to GaSe mixed with Ga$_2$Se$_3$ (blue), to pure phase Ga$_2$Se$_3$ (green). We summarize the XRD-determined phases as a function of growth temperature and Se/Ga atomic flux ratio in Fig. \ref{fig:phases}(d), where we find qualitative agreement with the phases expected from the Ellingham diagram. 

Increasing the growth temperature improves the standard crystal quality metrics, include a narrowing of the GaSe $004$ x-ray rocking curves indicating decreased mosaicity (Fig. \ref{fig:temperature}(a,b)), and a reduced surface roughness as measured by AFM (Fig. \ref{fig:temperature}(b,c-f)). These trends are in qualitative agreement with previous studies~\cite{Law2024}, although the observed rocking curve widths are broader than previous reports. GaSe growth on a smoothened GaAs buffer layer, rather than direct growth on de-oxidized GaAs (111)B, may be a route to decrease the mosaicity and further smoothen the resulting GaSe film surface. Buffer layers are a commonly used strategy for smoothening III-V and other surfaces for homoepitaxy~\cite{WHITWICK2008Buffer} and heteroepitaxy \cite{carlin2000impact} with reduced dislocation densities.

Note that for Se fluxes up to $2.25 \times 10^{15}$ atoms/cm$^2$s, GaSe film growth above a growth temperature of 475 $\degree$C leads to a sharp decrease of the Ga and Se sticking coefficients, and no crystalline growth is observed (Fig. \ref{fig:phases}(b,c) black curve). To access higher temperatures for improved atomic scale ordering, we employ a growth plus anneal approach, previously used by Ref.~\cite{Kis2018}, where we first grow a film at 400 $\degree$C where the sticking coefficient of GaSe is close to unity, then shutter the Ga and anneal the GaSe film at 520 $\degree$C under a combined Se$_\textrm{x}$ + Ga flux to balance against the rate of GaSe decomposition. We find that this high temperature anneal sharpens the x-ray rocking curve width and decreases the RMS surface roughness (Fig. \ref{fig:temperature}(b)). Minimal change in thickness was confirmed by XRR, indicating the observed improvements are structurally driven and not thickness-dependent.

\begin{figure} [t]
    \centering
    \includegraphics[width=1\linewidth]{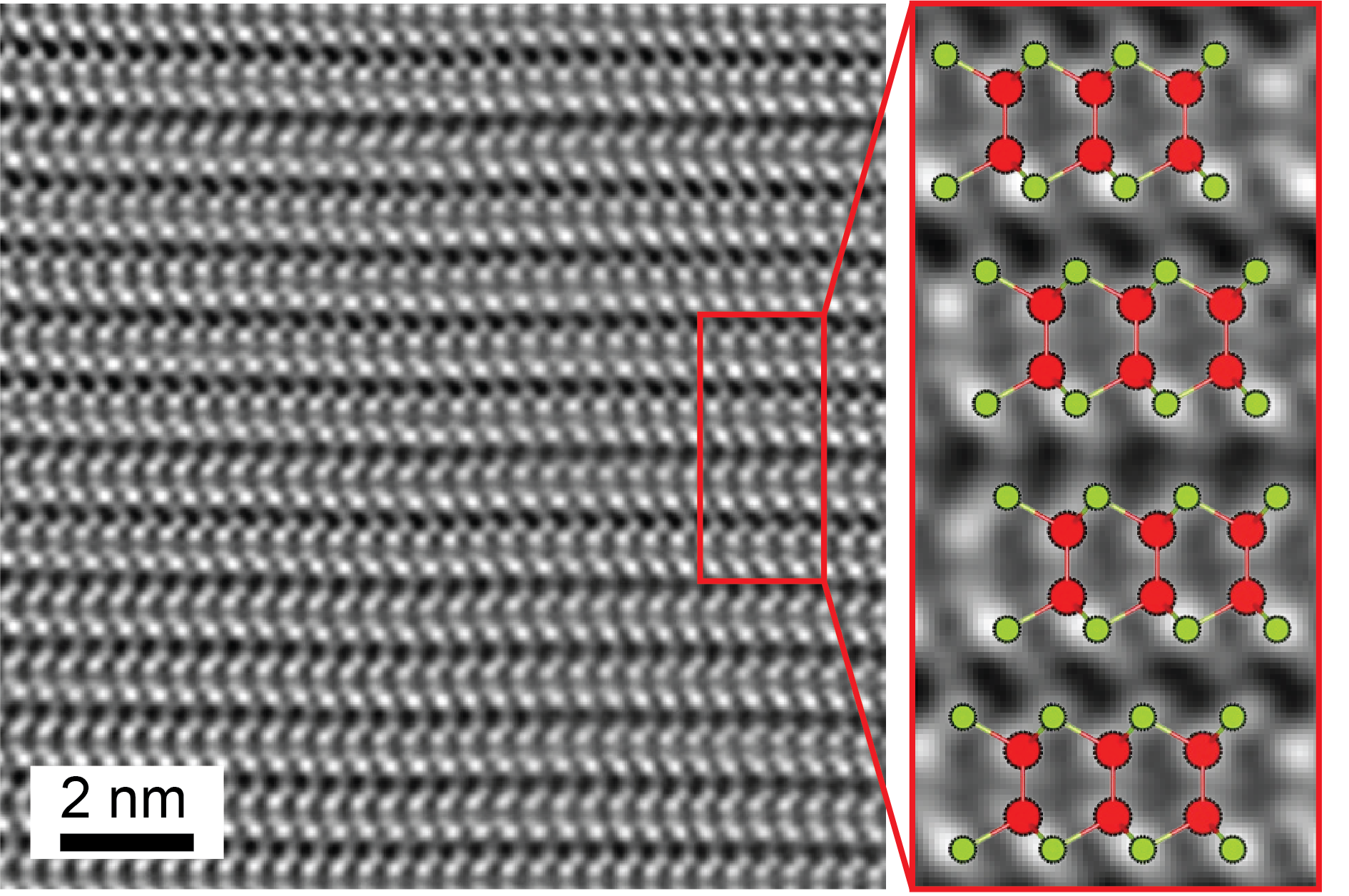}
    \caption{HAADF-STEM image of the GaSe film grown at 400 $\degree$C and annealed at 520 $\degree$C, measured along a $\langle100\rangle_\mathrm{GaSe} \parallel \langle110\rangle_\mathrm{GaAs}$ zone axis. Ga atoms are in red, and the Se atoms are in light green.
    }
    \label{fig:HAADF}
\end{figure}

Since it is difficult to distinguish the different GaSe polytypes and twinning by $00L$-type reflections alone, we use real space imaging by HAADF-STEM and off axis reflections by SAED and x-ray azimuthal scans. Figure \ref{fig:HAADF} shows a HAADF-STEM image of the film grown at 400 $\degree$C and annealed at 520 $\degree$C. The $ABC-ABC$ stacking with ``C'' shaped Se-Ga-Ga-Se tetralayers corresponds to the $\gamma$ polytype, and within this $\sim 180$ nm$^2$ field of view only a single orientation of $\gamma$ is imaged. 

\begin{figure*}[ht]
    \centering
    \includegraphics[width=1\linewidth]{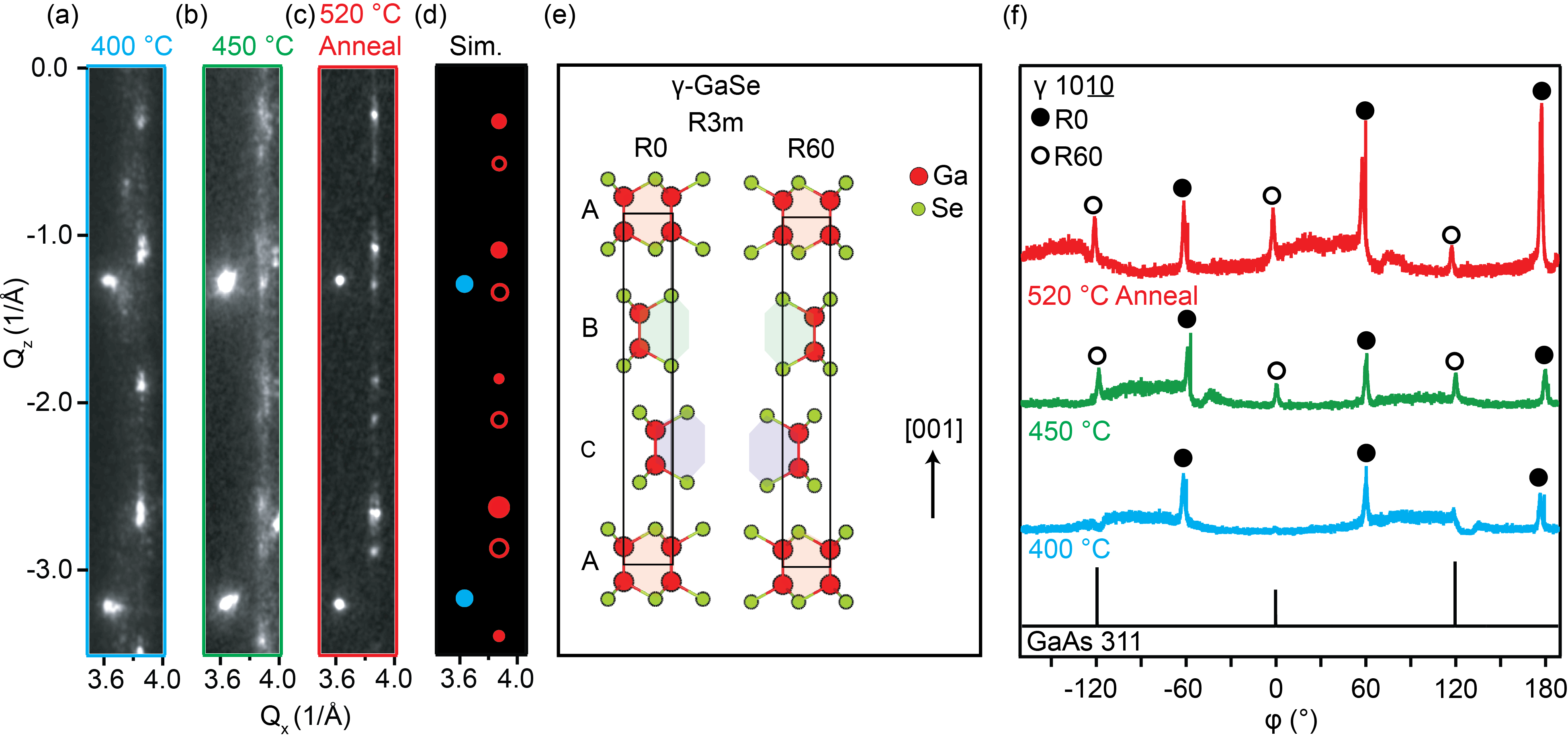}
    \caption{(a-c) Selected area electron diffraction (SAED) of the $20L$-type reflections for films grown at 400 $\degree$C and 450 $\degree$C, and the film annealed at 520 $\degree$C. Full SAED patterns in Appendix Figure \ref{fig:diffraction}. (d) Simulated electron diffraction pattern.
    (e) Crystal structure of $\gamma$-GaSe for the R0 and R30 twin orientations, viewed along $\langle 100 \rangle_\mathrm{GaSe}$ and $\langle 110 \rangle_\mathrm{GaSe}$, respectively. 
    (f) Azimuthal $\varphi$ x-ray diffraction scans of the GaSe $1 \, 0 \, \underline{10}$ film reflection, referenced to the GaAs substrate $311$ reflection.
    }
    \label{fig:polytype}
\end{figure*}

Selected area electron diffraction in the TEM, surveyed over a larger area (40 $\mu$m$^2$), indicate this annealed sample has R0 and R60 rotated $\gamma$ twin variants (Fig. \ref{fig:polytype}(c)). Here, twinning appears as double spots, since rhombohedral $\gamma$ ($R3m$) has 3-fold rather than 6-fold rotation around the $c$ axis. In Appendix Figure \ref{fig:linecut} we show that twinned $\gamma$ is the most consistent identification compared to the other potential polytypes.
X-ray diffraction azimuthal $\varphi$ scans over macroscopic areas (few mm$^2$) display an apparent 6-fold symmetry of the $\gamma$ $1\, 0\, \underline{10}$ (Fig. \ref{fig:polytype}(f), red curve), which arises from the superposition of R0 and R60 domains each with 3-fold rotation.

Interestingly, we find that while samples grown or annealed at  $T\geq 450$ $\degree$C are twinned, the sample grown at 400 $\degree$C is singly oriented. This is observed both in the SAED of the 400 $\degree$C sample, which appears as single rather than double spots (Fig. \ref{fig:polytype}(a)), and in the azimuthal x-ray scan, which shows 3-fold rather than 6-fold rotation (Fig. \ref{fig:polytype}(f), blue). 
We speculate that the appearance of singly-oriented $\gamma$ at lower growth temperature (400 $\degree$C), compared to twinned $\gamma$ at higher growth temperature (450 $\degree$C), arises due to thermal activation of mobile adatoms or ad-closers over kinetic barriers at high temperature. 

The effect of post growth annealing implies a slightly different mechanism. For the annealed film, presumably the initial GaSe growth at 400 $\degree$C produced singly oriented $\gamma$, and it is the act of annealing to 520 $\degree$C that produced twins. This suggests a solid-solid phase transition/reorientation of an already formed GaSe film, rather than differences in mobility of surface adatoms during film growth. Further studies are required in order to fully understand this 60 degree reorientation.

\section{Conclusion}

In summary, we mapped the adsorption-controlled MBE growth window for GaSe films on GaAs (111)B substrates and found a qualitative agreement with predictions from the Ellingham diagram. Increasing the growth temperature decreases the mosaicity and surface roughness, in agreement with previous reports. However, whereas the high temperature growth and annealing produce twinned $\gamma$ polytype, lower temperature growth at 400 $\degree$C produces singly oriented $\gamma$. Thus,  there is a compromise of desirable GaSe film metrics as a function of growth temperature. Further studies with more careful surface preparation, including smoothened GaAs buffer layers and well defined step edges, may be required to understand and better control twinning in MBE-grown GaSe films.



\section{Acknowledgments}

J.E. acknowledges support from the DEVCOM Army Research Laboratory (ARL) Research Associateship Program (RAP), administered by Oak Ridge Associated Universities (ORAU) for ARL through Cooperative Agreement (CA) W911NFT22T2T0097.

Additional support for MBE growth facilities (J.E. and J.K.K.) was provided by the NSF QLCI HQAN (NSF award No. 2016136). D.A.R. acknowledges additional support from the Office of Naval Research (N000142512276). 

We acknowledge the use of facilities and instrumentation in the Wisconsin Center for Nanoscale Technology. This Center is partially supported by the Wisconsin Materials Research Science and Engineering Center (NSF DMR-2309000) and by the University of Wisconsin–Madison.

\section{Author Declarations}

\subsection{Conflict of interest}
The authors have no conflicts to disclose.

\section{Data Availability}
\label{sec:DataAvail}

Raw data corresponding to each figure are made available at doi:xx to be finalized upon acceptance.

\clearpage \newpage \onecolumngrid 
\section{Appendix}

\begin{figure*}
    \centering
    \makebox[\textwidth][c]{%
        \includegraphics[height=0.4\textheight]{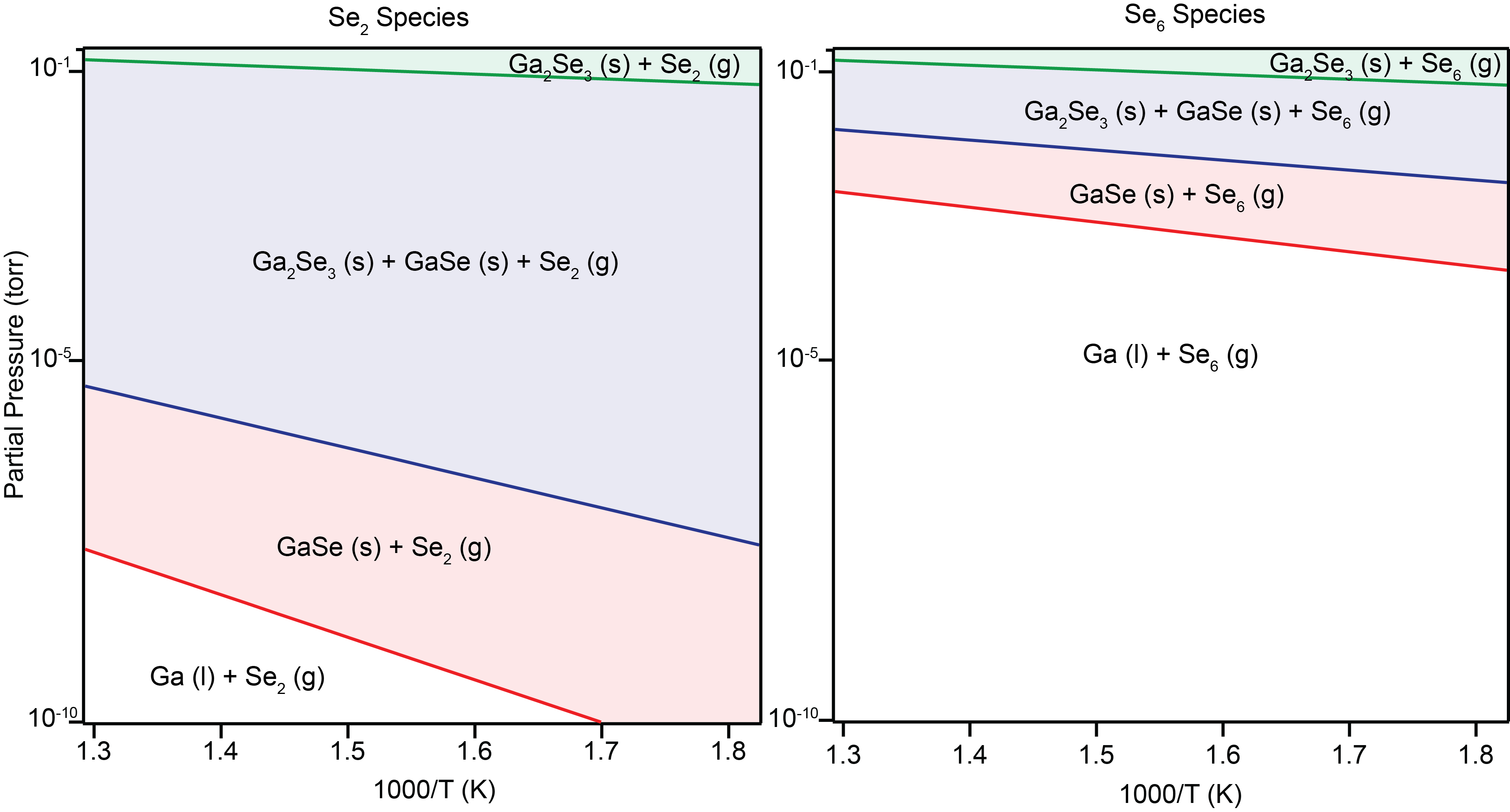}
    }
    \caption{Ellingham diagrams comparing Se$_2$ versus Se$_6$ as the gaseous species. \cite{Berkowitz1966,Gaur1981,Ider2015,Tyurin2005,SEDMIDUBSKY2019}
    }
    \label{fig:ellingham}
\end{figure*}

\begin{figure}[h]
    \centering
    \includegraphics[width=.8\linewidth]{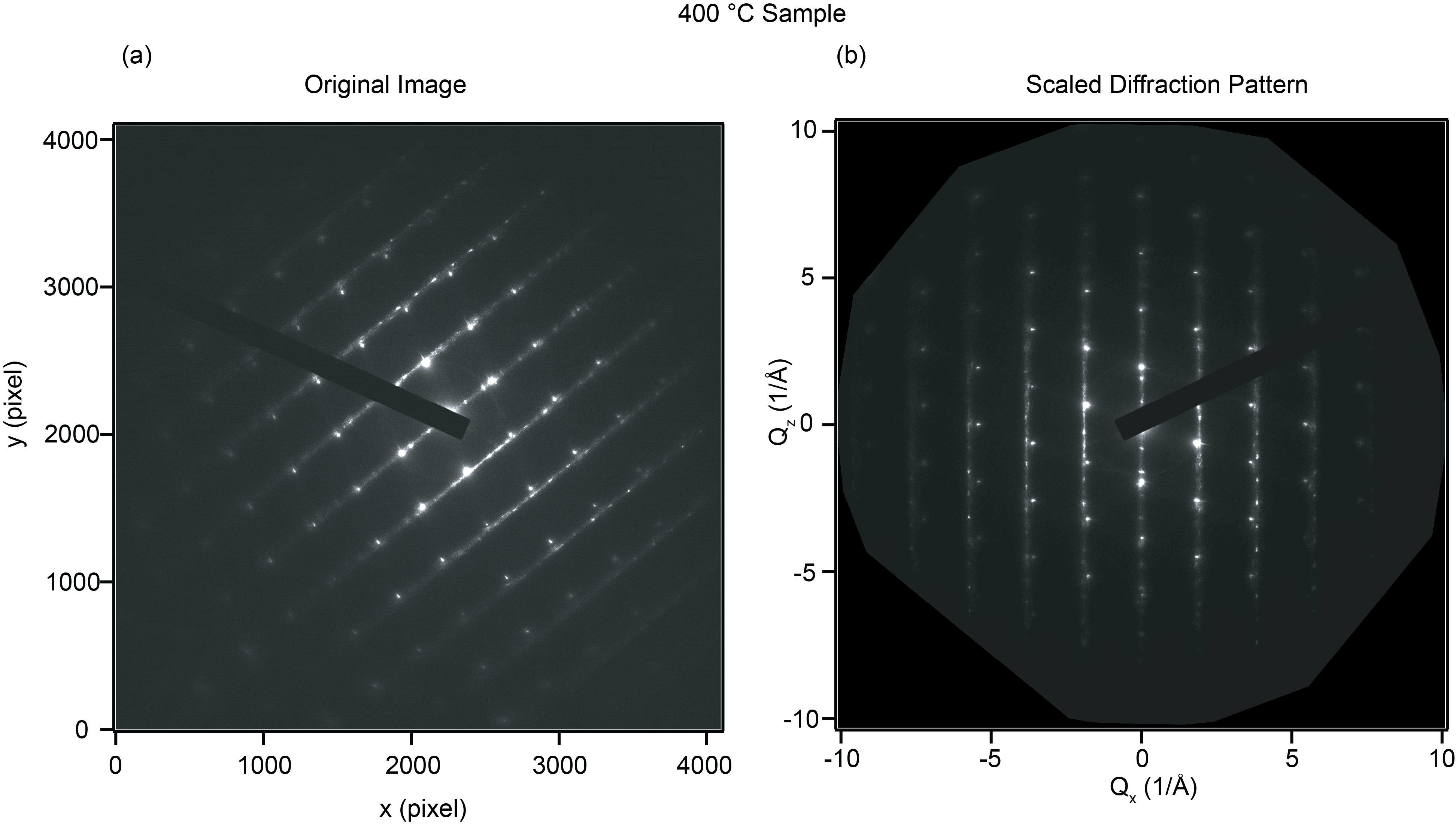}
    \caption{(a) Original SAED image for the 400 $\degree$C sample (b) Scaled diffraction pattern}
    \label{fig:Adjustment}
\end{figure}

\begin{figure*}
    \centering
    \includegraphics[width=1\linewidth]{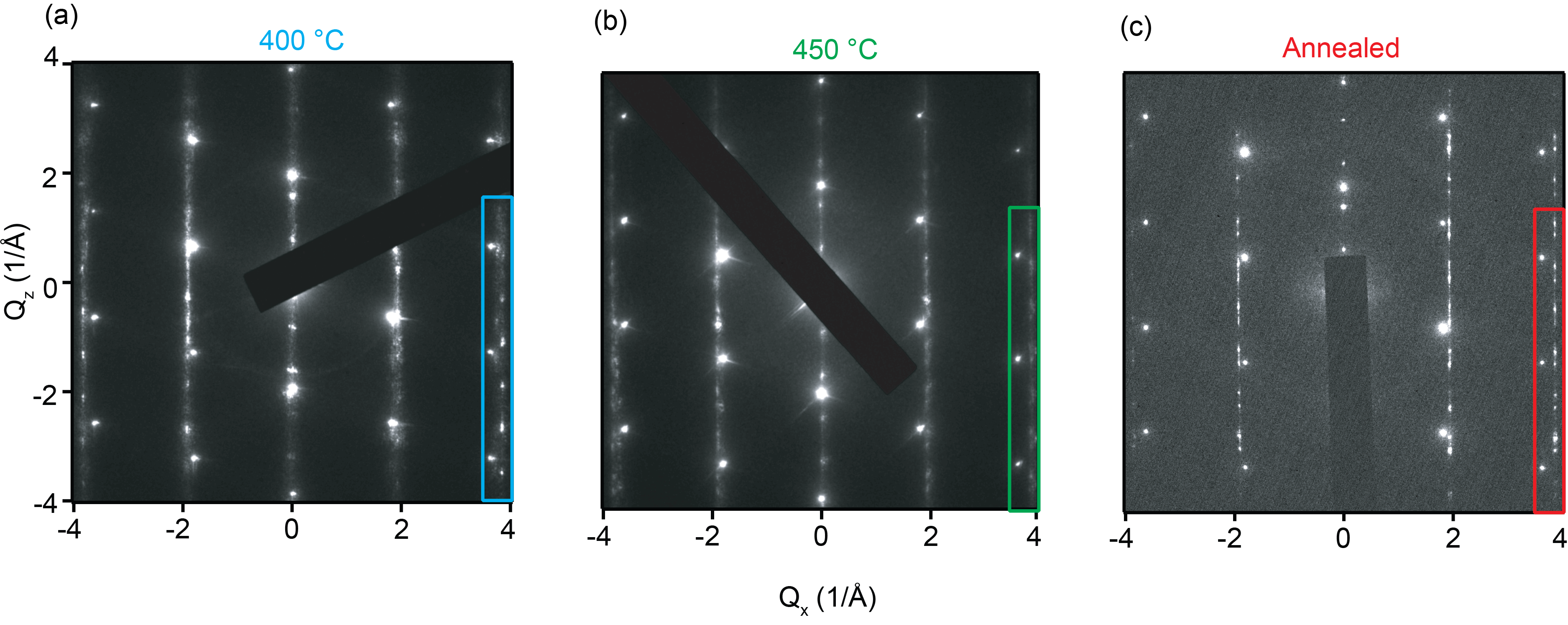}
    \caption{Electron diffraction patterns for (a) 400 $\degree$C, (b) 450 $\degree$C, and (c) 520 $\degree$C anneal samples. Boxes indicate locations of intensity line cuts.}
    \label{fig:diffraction}
\end{figure*}

\begin{figure*}
    \centering
    \includegraphics[width=1\linewidth]{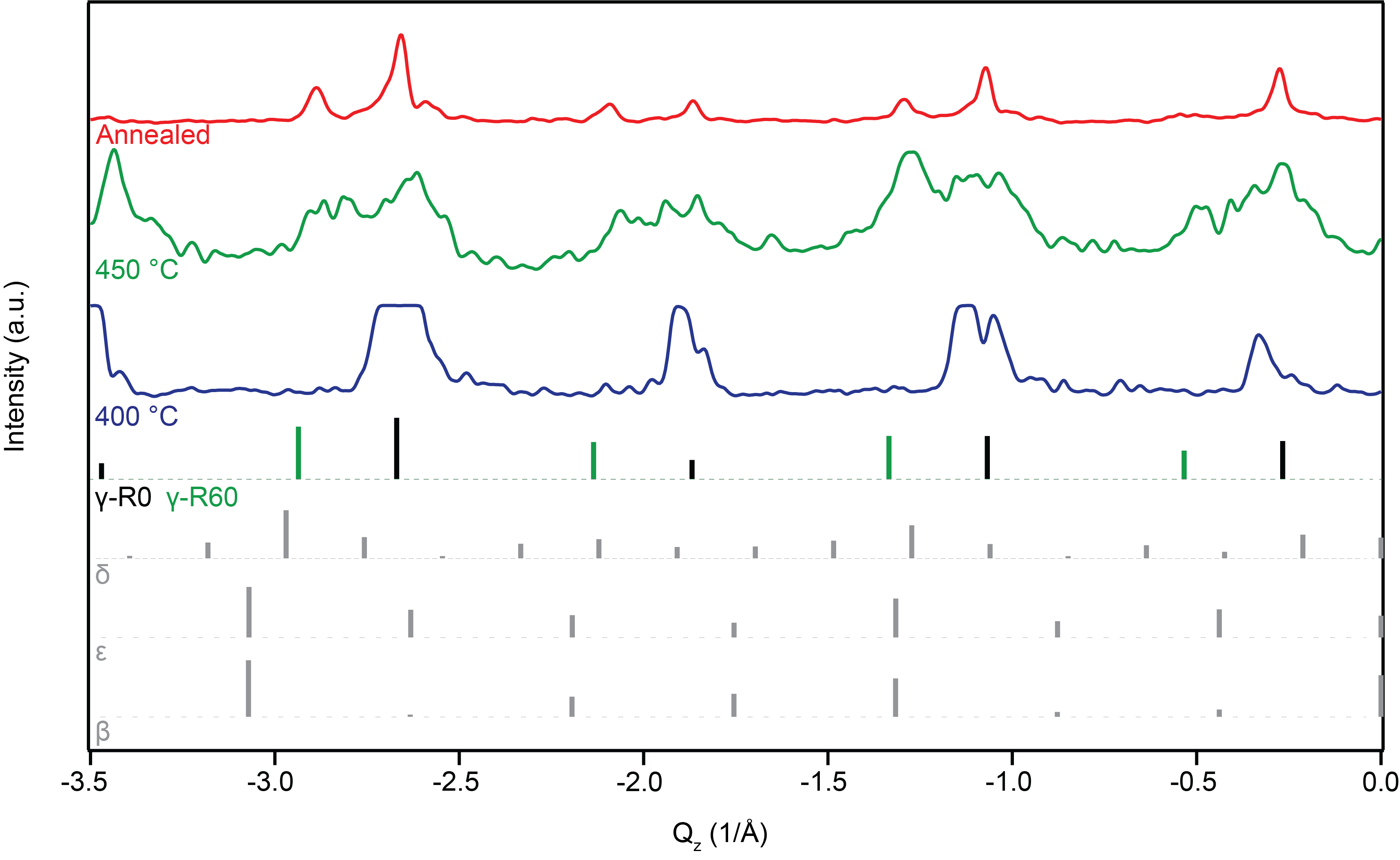}
    \caption{Intensity linecuts of the SAED at $H=2$, compared to simulated electron diffraction patterns for the different polytypes.}
    \label{fig:linecut}
\end{figure*}

\clearpage
\twocolumngrid

\bibliographystyle{unsrtnat} 
\bibliography{references}    

\end{document}